\documentclass{sig-alternate}
\usepackage{hyperref}
\hypersetup{colorlinks=true,citecolor=blue}
\usepackage{latexsym} 
\usepackage{hhline} 
\usepackage{comment}

\newcommand{\M}{\ensuremath{\mathsf{M}}} 
\newcommand{\I}{\ensuremath{\mathsf{I}}} 
 
\newcommand{\RR}{\mathcal{R}} 
 
\newcommand{\bigO}{\operatorname{\mathcal{O}}} 
\def\bigOsoft{\operatorname{\tilde{\mathcal{O}}}}

\newcommand{\Res}{\operatorname{Res}}
\newcommand{\val}{\operatorname{val}}
\newcommand{\Z}{{\mathbb{Z}}} 
 
\newcommand{\Q}{{\mathbb{Q}}} 
\newcommand{\N}{{\mathbb{N}}}

\newcommand{\Fp}{{\mathbb{F}_p}} 
 
\newtheorem{theorem}{Theorem} 
\newtheorem{proposition}{Proposition} 
\newtheorem{lemma}{Lemma} 
\newtheorem{definition}{Definition} 
\newtheorem{corollary}{Corollary} 
\newtheorem{example}{Example}

\begin{document} 
\conferenceinfo{ISSAC'06} {July 9--12, 2006, Genova, Italy.}
\CopyrightYear{2006} 
\crdata{1-59593-276-3/06/0007\\
This is the author's version of the work. It is posted here by permission of ACM for your personal use. Not for redistribution.} 

\title{Low Complexity Algorithms for Linear Recurrences} 
\subtitle{[Extended Abstract]}    

\numberofauthors{2} 
 
\author{ 
\alignauthor A. Bostan, F. Chyzak, B. Salvy \\ 
   \affaddr{Algorithms Project, Inria Rocquencourt}\\ 
   \affaddr{78153 Le Chesnay (France)}\\ 
   \email{{\{Alin.Bostan,Frederic.Chyzak,Thomas.Cluzeau,Bruno.Salvy\}@inria.fr}} 
\alignauthor T. Cluzeau\\ 
  \affaddr{Caf{\'e} Project, Inria Sophia Antipolis}\\ 
   \affaddr{06902 Sophia Antipolis (France)}\\ 
}

\maketitle

\begin{abstract} 
We consider two kinds of problems: the computation of polynomial and rational solutions of linear recurrences with coefficients that are polynomials with integer coefficients; indefinite and definite summation of sequences that are hypergeometric over the rational numbers.
The algorithms for these tasks all involve as an intermediate quantity an integer~$N$ (dispersion or root of an indicial polynomial) that is potentially exponential in the bit size of their input. Previous algorithms have a bit complexity that is at least quadratic in~$N$. We revisit them and propose variants that exploit the structure of solutions and avoid expanding polynomials of degree~$N$. We give two algorithms: a probabilistic one that detects the existence or absence of nonzero polynomial and rational solutions in~$\bigO(\sqrt{N}\log^{2}N)$ bit operations; a deterministic one that computes a compact representation of the solution in~$\bigO(N\log^{3}N)$ bit operations. Similar speed-ups are obtained in indefinite and definite hypergeometric summation. We describe the results of an implementation.
\end{abstract} 

\bigskip
\noindent
{\bf Categories and Subject Descriptors:} I.1.2 {[{\bf Symbolic and
    Algebraic Manipulation}]}: {Algorithms}
    
    \vspace{1mm}
\noindent
{\bf General Terms:} Algorithms, Experimentation, Theory

\vspace{1mm}
\noindent
{\bf Keywords:} Computer algebra, polynomial and rational solutions,
 linear recurrences, summation, creative telescoping, complexity

\section{Introduction}  
A central quantity for many algorithms operating on linear recurrences and their solutions is the dispersion.
\begin{definition}
The \emph{dispersion set} of two polynomials $P$ and $Q$ in $\Q[n]$ is the set of positive integer roots 
of the resultant $R(h)=\Res_n(P(n),Q(n+h))$. When this set is not empty, its maximal element is called the \emph{dispersion} of $P$ and~$Q$.
\end{definition}
Thus, the dispersion is the largest integer difference between roots of~$P$ and~$Q$.
As shown by the simple example~$(P,Q)=(n,n-N)$ with~$N\in\N$, the dispersion can be exponentially large in the \emph{bit} size of the input polynomials. It cannot get much worse: when the polynomials have integer coefficients whose absolute value is bounded by~$B$, their dispersion is bounded by $4B$ \cite[Fact~7.11]{Gerhard04}. 
This exponential size yields the dominant term in the worst-case complexity of many algorithms computing~---~or operating on~---~solutions of linear recurrences.

For instance,  the computation of a Gosper-Petkov\v{s}ek form produces a polynomial whose degree~$N$ can be linear in the dispersion of its input and thus exponential in its bit size. If this polynomial is expanded it has $N+1$ coefficients; over the integers, its total bit size is~$\bigO(N^2\log N)$. This form is used in the first step of Gosper's summation algorithm and of Abramov's algorithm for computing rational solutions of linear recurrences. Thus, it makes an important contribution to the complexity of these algorithms. Once this form is computed, these algorithms search for polynomial solutions of an associated linear recurrence. This is done by linear algebra using a bound on the possible degree of solutions which is at least as large as~$N$, leading again to a more than quadratic complexity, even when no nonzero solution exists. In turn, a parameterized variant of Gosper's algorithm forms the basis of Zeilberger's definite summation algorithm which inherits this costly behaviour. By contrast, we provide a probabilistic algorithm that detects that no nonzero rational solution of a homogeneous linear recurrence exists in~$\bigO(\sqrt{N}\log^2N)$ bit operations and a deterministic algorithm that gives a compact representation of all solutions in~$\bigO(N\log^3N)$ bit operations.
All the algorithms in the present work eventually rely on the computation of polynomial solutions of linear recurrences. In a previous work~\cite{BoClSa05}, we dealt with the analogous problem in the linear differential case, by exploiting the linear recurrence satisfied by the coefficients of power series solutions and reducing the computation to that of \emph{matrix factorials}. For the latter operation, there exist fast probabilistic and deterministic algorithms~(see~\cite{BoGaSc03,ChCh88} and the references in~\cite{BoClSa05}). In the case of linear recurrences, it is not true that the coefficients of polynomial solutions satisfy a linear recurrence in general; however, it becomes true if the polynomials are expanded in a binomial basis~\cite[Ch.~XIII, art.~5]{Boole1872}. This is the basis of a simple quadratic algorithm~\cite{AbBrPe95} to compute polynomial solutions. In Section~\ref{sec1}, we show how this conversion is performed, we recall the basic results on matrix factorials and apply them to get the announced complexities.

From there, in Section~\ref{sec2}, we proceed in three steps: {\em (i)\/}~we slightly modify the computation of the Gosper-Petkov\v{s}ek form so that it does not expand the potentially large polynomial but instead computes a first-order, moderately-sized recurrence for it; {\em (ii)\/}~we show that this first-order recurrence can be used to compute a linear recurrence satisfied by the numerators of rational solutions, in a complexity that is only logarithmic in~$N$, both in the homogeneous and nonhomogeneous cases; {\em (iii)\/}~we then compute the numerators as polynomial solutions via matrix factorials. The close relation between Abramov's and Gosper's algorithm makes it possible to transfer these results to Gosper's algorithm. Then in Section~\ref{zeilberger}, we show how this machinery can be adapted to the parameterized variant needed in Zeilberger's algorithm.
Finally, we describe experimental results in Section~\ref{experiments}.
 
\medskip {\bf {\noindent Notations and complexity measures.}}  All
along this text, $\RR$ denotes a linear difference operator with
coefficients in~$\Z[n]$.  We view it as a polynomial in the non-commutative ring $\Q\langle n,S_n\rangle$, where $S_n$ is the shift operator $S_n u(n) = u(n+1)$. Similarly, $S_x$, $S_k$, and~$S_m$ denote the shifts with respect to $x$, $k$, and~$m$. To any difference operator $\RR$ is attached a 
homogeneous linear recurrence equation $\RR u = 0$. We view the
solution $u$ either as a sequence $(u_n)$ (also denoted $u_n$), or as
a function $u(n)$ (the cases of particular interest being polynomial
and rational functions).

For our complexity analyses, the measure we use is the bit (or
boolean) complexity. For this purpose, our complexity model is the
multi-tape Turing machine, see for instance~\cite{ScGrVe94}. We use
the \emph{number of bit operations} to express time complexities in this model.
We call \emph{bit size} (or simply~\emph{size}) of an integer~$a
\neq 0$ the number~$\lambda(a):=\lfloor\log|a| \rfloor+1$ ($\log x$ denotes
the logarithm of~$x$ in base~2). By convention, we assume
that~$\lambda(0)=1$. The bit size of a matrix or vector is the sum of the bit sizes of its entries. Polynomials given as input to our algorithms are stored in a dense representation; a measure of their bit size is given by the sum of the bit sizes of their coefficients, including the zero coefficients. Similarly, the bit size of a linear recurrence equation (LRE) is the sum of the bit sizes of its polynomial coefficients.

To simplify complexity estimates, we assume that the product of two
integers of bit size~$d$ can be computed within $\I(d) = \bigO (d \log
d \log\log d)$ bit operations using Fast Fourier Transform~\cite{ScSt71}.  To keep
the notation compact, we sometimes write $\I(d) = \bigOsoft (d)$; the tilde
indicates that the factors polynomial in $\log d$ or smaller have been omitted.

For any prime number~$p$, the bit complexities of the operations
$(+,-,\times ,{\div})$ in the finite field $\Fp:=\Z/p\Z$ are in $\bigO(\I(\log p)\log\log p)$.  We assume that over the rings we use, the product of two
polynomials of degree at most~$d$ can be computed within~$\bigO
(\M(d))$ base ring operations (each ring operation being counted at
unit cost) and that $\M(d)=\bigOsoft(d)$~\cite[ch.~2]{BuClSh97}.  For computations in $\Fp[x]$, the bit complexity is
bounded by multiplying the arithmetic cost estimates by the bit
complexity of the basic operations in~$\Fp$.

In all our algorithms we are interested in reducing the complexity with respect to a potentially exponential parameter~$x$ (related to a dispersion or to a root of an indicial polynomial). Thus we consider as having cost~$\bigO(1)$ any operation whose complexity is polynomial in the bit size of the input recurrence or polynomials, and concentrate on the dependency of the complexity in~$x$. In order to provide the code with an actual bound on the size of primes that need to be used so that the bound on probability of error is guaranteed, we have to perform a much more precise complexity analysis taking into account all parameters (order, degree of coefficients) (as in the proof of~\cite[Thm.~3]{BoClSa05}). Such a detailed analysis will appear in~\cite{BoChClSa06}.

\section{Polynomial Solutions} \label{sec1}

In symbolic summation and in the resolution of linear recurrences, all the
known algorithms ultimately require polynomial solutions
of linear recurrence equations.

In this section, we give algorithms for computing descriptions of the $\Q$-vector space of solutions of a linear
recurrence operator $\RR$ with coefficients in $\Z[n]$:
\begin{equation} \label{rec} \RR u  = 
  a_{r}(n) u(n+r) + \cdots + a_0(n) u(n)=0, \quad  n
  \geq 0.
\end{equation}
We focus on two types
of solutions of such recurrences: solutions with finite
support and polynomial solutions.

In what follows, we make the hypothesis that \emph{$0$ is an ordinary
  point of the recurrence}. This means that the leading coefficient
$a_r(n)$ does not vanish at any of the integers $0,1,2,\ldots$; in
other words, when unwinding the recurrence, no division by zero is
encountered. This condition is ensured after a generic translation $n
\mapsto n+\alpha$. Under our complexity assumptions, a proper~$\alpha$ and the corresponding translation can be computed in a polynomial number of bit operations, so that there is no loss of generality for the problems we consider. The general case (when 0 is not ordinary) is technically more demanding but does not change the complexity estimates we give here. It will be presented in~\cite{BoChClSa06}.

Let $\operatorname{Sol}(\mathcal{R})$ denote the vector space of
solutions $(u_0,u_1,\ldots)$ of~\eqref{rec}. In the case of an arbitrary $\mathcal{R}$, the dimension of
$\operatorname{Sol}(\mathcal{R})$ as a $\Q$-vector space may be different from
$r$ (both larger or smaller).  However, when~0 is an ordinary point of~$\RR$, 
$\operatorname{Sol}(\mathcal{R})$ has dimension exactly $r$ and a basis is given by the sequences~$u^{(j)}$, $j=0,\dots,r-1$, satisfying~$\RR$ and having initial conditions $u_i^{(j)}=(\delta_{j,i})_i$, for $0\leq i \leq r-1$, where $\delta_{m,n}$ is Kronecker's $\delta$ symbol ($\delta_{m,m}=1$, $\delta_{m,n}=0$ if $m\neq n$).

In \S\ref{compact}, we describe the  compact representation that forms the basic data structure of our algorithms. Then, in \S\ref{Nthterm} we recall classical results that allow for  the efficient
computation of the $N$th element of a solution of~$\RR$.
In~\S\ref{rectorec} we describe the reduction from the
problem of searching for polynomial solutions to that of finding
solutions with finite support. Next, we give in~\S\ref{polysols}
algorithms to compute finitely supported and polynomial solutions of
recurrences. We conclude this section by showing in
\S\ref{compactoperations} how the evaluation of a polynomial and its finite
differences can be performed efficiently in the compact representation.

\subsection{Compact Representation}\label{compact}
Classically, a polynomial solution~$u(n)$ of~\eqref{rec} is represented by its
coefficients in the monomial basis $\{n^k\}$.  We use an alternative
data structure for $u(n)$, which is motivated by the observation that
its coefficients $c_k$ in the binomial basis $\{\binom{n}{k}\}$
obey a recurrence with polynomial coefficients.

\begin{example}\label{ex:compact} The recurrence $(n+1)
  u(n+1) - (n+N+1) u(n)=0$ has a unique nontrivial monic polynomial
  solution $u(n) = (n+1) \dotsm (n+N)$.  To write down its
  coefficients in the monomial basis at least $\frac{1}{2}N^2 \log N$ bits are
  needed. In contrast, $u(n)$ can be represented by the recurrence
  \[(k+1) c_{k+1} -(N-k) c_k=0,\quad c_0 = N!\] on
  the coefficients $c_k$ of $u(n)$ in the binomial basis
  $\{\binom{n}{k}\}$; the bit size of this new representation is
  only linear in $N\log N$, most of the size being in the initial condition~$c_0=N!$.
\end{example}
\begin{definition}The \emph{compact representation} of a polynomial solution of~\eqref{rec} is the data of a linear recurrence and initial conditions for its coefficients in the binomial basis, together with an upper bound on its degree.
\end{definition}
Our aim in this article is to demonstrate that this representation of polynomial solutions of recurrences can be
 carried through different algorithms from indefinite and definite
 hypergeometric summation and that it is beneficial from the
 complexity point of view.
The reason why this representation deserves the name ``compact'' appears in~\S\ref{polysols} below.

\subsection{High-Order Terms of Sequences}\label{Nthterm} 
Let $(u_n)$ be a sequence satisfying~\eqref{rec}. The recurrence~$\mathcal{R}$ can be rewritten as a first-order matrix recurrence $U_{n+1} =
\mathcal{C}(n+1) U_n$, where $U_n$ is the vector~$(u_n,u_{n-1},\dots,u_{n-r+1})^t$ and
$\mathcal{C}$ is an $r \times r$ matrix
with rational function entries. The problem of computing a selected
term $u_N$ reduces to that of computing~$U_{r-1}$ and the \emph{matrix factorial}
$\mathcal{F}(N):=\mathcal{C}(N) \cdots \mathcal{C}(r)$.  This makes
sense since under our hypothesis the leading term of the initial recurrence does not
vanish at $1,2,\ldots, N$. The numerator and denominator of the matrix factorial can be computed
efficiently, either in $\Z$ using a \emph{binary splitting}
algorithm, or modulo a prime $p$ using a \emph{baby-step/giant-step} algorithm. These algorithms are described in~\cite[\S 2.1, \S
3.1]{BoClSa05}, see the references therein. For further use, we
extract from~\cite{BoClSa05} the following result.

\begin{theorem}[\cite{BoClSa05}] \label{thm:reccomplexity} Let $(U_i)$ be a sequence of
  vectors of rational numbers that satisfies a recurrence $U_{i+1}=C(i+1)U_i$, with $C(x)$ an $r\times r$ matrix with rational function entries in $\Q(x)$. 
Let $p$ be a prime number
  such that the denominator of~$C$ does not vanish mod 
  $p$ at $1,2,\ldots, N$. Then, as~$N\rightarrow\infty$:
\begin{enumerate}
\item[(a)] $\mathcal{F}(N)=C(N)\dotsm C(r)$ and $U_N$ have bit size $\bigO(N \log N)$; their values can be
  computed using $\bigO \big( \I (N \log N) \log N \big)$ bit operations.
\item[(b)] $\mathcal{F}(N)\bmod p$ and $U_N\bmod p$ can be computed using\\ $ \bigO
  \left( \M(\sqrt{N})\I (\log p )\right)$ bit operations.
\item[(c)] The rank of $\mathcal{F}(N)$ can be computed in~$\bigOsoft(\sqrt{N})$ bit operations using a probabilistic Monte Carlo algorithm.
\end{enumerate}
\end{theorem}

\subsection{Expansion in the Binomial Basis} \label{rectorec}
For completeness, we recall here an algorithm from~\cite{AbPeRy00} that we call \textsf{RecToRec} to perform the
conversion from a recurrence with polynomial coefficients to the
recurrence satisfied by the coefficients of series solutions in
the binomial basis. Earlier (and slightly more complicated) algorithms have been given in~\cite[Chapter~XIII]{Boole1872} and~\cite[Section~4.2]{AbBrPe95}.
The starting point are the following two identities:
\[\binom{n+1}{k}=\binom{n}{k}+\binom{n}{k-1},\ 
n\binom{n}{k}=k\binom{n}{k}+(k+1)\binom{n}{k+1}.\]
If $u(n)=\sum_{k=0}^\infty{c_k\binom{n}{k}}$, then applying these identities to rewrite~$u(n+1)$ and~$nu(n)$ and extracting coefficients of~$\binom{n}{k}$ shows that the ring morphism $\phi:\Q[n,S_n]\rightarrow\Q[k,S_k,S_k^{-1}]$ defined by~$\phi(S_n)=1+S_k$ and~$\phi(n)=k(1+S_k^{-1})$  sends a homogeneous LRE satisfied by~$u(n)$ to another one satisfied by~$c_k$. The image of~\eqref{rec} is a LRE of the form
\[
(a_r(k)S_k^r+b_{r-1}(k)S_k^{r-1}+\dots+b_{-s}(k)S_k^{-s})c_k=0,\quad k\ge0,
\]
where the leading term is exactly that of~\eqref{rec} and the trailing term may involve a negative shift (by convention, $c_k=0$ when $k<0$). In
particular, if $0$ is an ordinary point for $\mathcal{R}$, so is it for $\phi(\mathcal{R})$.
The resulting algorithm is as follows. Its complexity is clearly polynomial in the bit size of~$\mathcal{R}$. 
\begin{center}
\fbox{
\begin{minipage}{8cm}
\begin{center}
\textsf{Algorithm RecToRec}\\
\end{center}
{\bf Input}: a recurrence  $\mathcal{R} u = 0$, where $u(n) =
\sum_k c_k \binom{n}{k}$. \\ 
{\bf Output}: a recurrence $\mathcal{S}$ satisfied by the sequence $(c_k)$, plus a set~$\mathcal{E}$ of linear equations on its initial conditions. 
\begin{enumerate}
	\item Compute~$\mathcal{T}=\phi(\mathcal{R})$;
	\item Let $-s=\val_{S_k}(\mathcal{T})$ be its valuation w.r.t.~$S_k$;
	\item If~$s<0$ return $\mathcal{S}:=\mathcal{T}$ and $\mathcal{E}:=\emptyset$,
	\item Otherwise return $\mathcal{S}:=S_k^s\mathcal{T}$ and the equations $\mathcal{E}:=\{\left.((S_k^i\mathcal{T})c)\right|_{k=0}=0, i=0,\dots,s-1\}$.
\end{enumerate} 
\end{minipage}
}
\end{center}

\subsection{Finite Support and Polynomial Solutions}\label{polysols}
We consider here the problem of computing a basis of solutions
\emph{with finite support}, that is, whose terms beyond a certain
index are all zero. The \emph{degree} of a solution with finite support $u$
is, by definition, the unique integer $n$ such that $u_n \neq 0$ and
$u_{n+i}=0$, for all $i \geq 1.$ A universal bound $N$ on the degrees
of all solutions with finite support of the input recurrence
$\mathcal{R}u=0$ is given by the largest positive integer root of the
trailing coefficient $a_0(n)$ of $\mathcal{R}$.  Note that $N$ is
generally not bounded polynomially in the bit size of~$\mathcal{R}$.

Recall that $\operatorname{Sol}(\mathcal{R})$ has dimension $r$, with
a basis~$\mathcal{B}$ formed by the sequences
$u^{(j)}$, with $0\leq j \leq r-1$, given by the initial conditions
$u^{(j)}_i = \delta_{j,i}$, for $0\leq i \leq r-1$.  Thus, a finitely
supported solution $u$ is (an unknown) linear combination
$\sum_{j=0}^{r-1} \lambda_j u^{(j)}$ such that the elements in the
slice $u_{N+1},\ldots, u_{N+r}$ all vanish. This yields linear constraints on the initial
conditions $\lambda_j$.

To determine these constraints, it is sufficient to compute the values at
indices ${N+1},\ldots, {N+r}$ of all the elements in $\mathcal{B}$
using Thm.~\ref{thm:reccomplexity}. The rank of the
resulting $r \times r$ matrix   gives the dimension of the vector space of solutions
with finite support.  Since the entries of this matrix have bit size
$\bigO(N \log N)$, the desired $\lambda_j$'s, which are determined by a kernel
computation, also have bit size $\bigO(N \log N)$. Putting
together these considerations, we get the following result.

\begin{theorem}\label{th:finitesuppsols}
There exists a basis~$(u^{(1)},\dots,u^{(d)})$ ($d\leq r$) of solutions of
recurrence~(\ref{rec}) with finite support, where each~$u^{(i)}$ is uniquely specified by
the data of {initial conditions} of bit size $\bigO(N \log N)$,
with $N$ a bound on the integer roots of $a_0$.  The
dimension~$d$ as well as the degrees of the~$u^{(i)}$'s
can be computed by a probabilistic algorithm using $\bigOsoft( \M(\sqrt{N})  \I (\log N )) $ bit operations.  The
initial conditions of the~$u^{(i)}$'s and their maximal degree $D
\leq N$  can be computed  deterministically
in $\bigO(\I(D \log D) \log D)$ bit operations.
\end{theorem}
Thm.~\ref{th:finitesuppsols} is the basis for using the name ``compact representation'': it shows that the compact representation has a size of the same order as the initial conditions, while the expanded polynomials have size~$\bigO(N^2\log N)$. In general, this latter bound is reached. 

Using the results in \S\ref{rectorec}, 
Thm.~\ref{th:finitesuppsols} carries over literally to the compact
representation of a basis of polynomial solutions of the recurrence
$\mathcal{R}u=0$. The corresponding statement requires a bound on the degree of polynomial solutions that is given by the roots of the \emph{indicial polynomial}. 
\begin{definition}The \emph{indicial polynomial} of~$\mathcal{R}$ at infinity is the trailing coefficient of\/ $\sf{RecToRec}(\mathcal{R})$.
\end{definition}
\begin{corollary}\label{cor:finitesuppsols} The statement of Thm.~\ref{th:finitesuppsols} holds for polynomial solutions of~$\mathcal{R}u=0$, with~$N$ the largest integer root of the indicial polynomial of~$\mathcal{R}$ at infinity.
\end{corollary}

\medskip \noindent {\bf Nonhomogeneous Equations.}
We now consider the equation $\mathcal{R}u(n) = f(n)$, with coefficients in~$\Z[n]$ and
right-hand side of degree $m$. 
Applying
\textsf{RecToRec} and expanding $f(n)$ in the binomial basis, the
initial problem boils down to the search of finitely supported solutions
of a nonhomogeneous equation $\mathcal{S}c(k) = g(k)$, where $g$ is a sequence with finite support, $g(i) = 0$ for $i >
m$. In matrix notation, we have $U_{k+1} = C(k+1) U_k +
v_{k+1}$, where $U_k$ is the vector $(u_k,\ldots, u_{k-r+1})^t$ and $v_k$
is the vector $(g(k),0,\dots, 0)^t$. Then the vector of initial
conditions~$U_{r-1}$ satisfies the affine constraint $A (BU_{r-1} + w_{m}) =
0$, where $A:=C(N+r) C(N+r-1) \cdots C(m+1)$, $B:=C(m)C(m-1) \cdots C(r)$,
$w_i:=v_i + C(i) w_{i-1}$, for $r+1\leq i \leq m+1$ and $w_r=v_r$.

For large $N$, using Thm.~\ref{thm:reccomplexity}, the matrices $A$ and $B$ can be
computed efficiently.   
The
bit size and the computational cost of $w_{m+1}$ is $\bigO(1)$. Thus,
solving the affine system of size $\bigOsoft(N)$ yields the finitely
supported solutions of $\mathcal{S}c=g$ and the polynomial solutions of
$\mathcal{R}u=f$ and we get the following.
\begin{corollary} Let $f$ be a polynomial. Then the statement of~Thm.~\ref{th:finitesuppsols}
holds for polynomial solutions of nonhomogeneous equations~$\mathcal{R}u(n)=f(n)$ as
the largest integer root~$N$ of the indicial polynomial of~$\mathcal{R}$ at infinity becomes large.
\end{corollary}

\subsection{Evaluation in Compact Representation}\label{compactoperations}
The compact representation is not only a data structure for intermediate computations. It can actually be exploited further. In particular, we now detail the evaluation at an algebraic number $\alpha$ 
of a
polynomial $u(x)$ and an iterated difference $\Delta^{H}(u)$ (where~$\Delta=S_x-1$ and $H$ is potentially large).
The polynomial $u$ is given by its degree $N$ and 
the recurrence
$$\sum_{i=0}^r a_{i}(k) c(k+i)=0 \quad \text{for all}
\; k \geq 0$$
satisfied by its coefficients $c_k$ in the
binomial basis $\{\binom{x}{k}\}$, together with initial conditions.
The basic idea is embodied in the following.
\begin{lemma}[Folklore]If $(u_k)$ and $(v_k)$ are solutions of linear difference equations with polynomial coefficients,
	then so is the sequence $(u_N)$ defined by $u_N=\sum_{k=0}^N{u_kv_k}$.
\end{lemma}
This lemma can be applied to the sequences $(c_k)$ and $\binom{\alpha}{k}$. Evaluating the resulting sequence
at~$N$ using Thm.~\ref{thm:reccomplexity} gives $u(\alpha)$ for $\bigO(\I(N \log N) \log N)$ bit operations, when $N$ is large.

Using Pascal's formula $\Delta^H \binom{x}{i} = \binom{x}{i-H}$, we
deduce that $\Delta^Hu(\alpha) = \sum_{k=0}^{N-H} c_{k+H} \binom{\alpha}{k}.$ The recurrence satisfied by the sequence $(c_{k+H})_k$
is obtained by shifting by~$H$ the coefficients of the recurrence of
$(c_k)$. This new recurrence has bit size $\bigO(\log H)$ and initial
conditions can be determined by binary splitting in $\bigO(\I(N \log N
\log H) \log N \log H)$ bit operations. Here, our asymptotic bound involves the two parameters $N$ and~$H$, as both are potentially exponential in the input size.
As above, the
compact representation of the recurrence satisfied by $D_k :=
\sum_{\ell=0}^{k} c_{\ell+H} \binom{\alpha}{\ell}$ can be determined
efficiently, as well as its $N$th term 
$\Delta^Hu(\alpha)$.

\section{Rational Solutions}\label{sec2}
\subsection{Compact Gosper-Petkov\v{s}ek Normal Form}
The classical Gosper-Petkov\v{s}ek normal form~\cite{Petkovsek92,Gosper78} of a reduced rational function~$P/Q$ in $\Q(n)$ consists of three polynomials $A,B,C$ in 
$\Q[n]$ such that 
\begin{equation} \label{gp}
\frac{P(n)}{Q(n)} = \frac{A(n)}{B(n)} \frac{C(n+1)}{C(n)},
\end{equation}
with the constraints
\begin{multline} \label{gp-ext}
\gcd(A(n),C(n))=1, \ \gcd(B(n),C(n+1))=1,\\
\text{and for all $h\in\mathbb{N}$,} \ \gcd(A(n),B(n+h))=1.
\end{multline}
The degree~$N$ of the polynomial~$C(n)$ 
is potentially exponentially large. Thus, in our algorithm \textsf{CompactGPF} below, we modify  the usual algorithm (e.g., in~\cite{PeWiZe96}) slightly so that the polynomial~$C(n)$ is not expanded. Similar ideas appear in~\cite{GeGiStZi03} in the context of indefinite rational summation.
\begin{center}
\fbox{
\begin{minipage}{8cm}
\begin{center}
\textsf{Algorithm CompactGPF}\\
\end{center}
{\bf Input:} an ordered pair $(P(n),Q(n))$ of polynomials.\\
{\bf Output:} $(A(n),B(n),\{(g_i(n),h_{i}),i=1,\dots,s\})$ such that $C(n)=\prod_{i}{g_{i}(n-1)\dotsm g_{i}(n-h_{i})}$ satisfies (\ref{gp}).
\begin{enumerate}
\item Compute $h_1>\dots>h_s>0$ the positive integer roots of $\Res_n(P(n),Q(n+h))$; 
\item $A(n):=P(n), B(n):=Q(n)$;
\item For $i$ from $1$ to $s$ do
\item[a.] $g_i(n):=\gcd(A(n),B(n+h_i))$;
\item[b.] $A(n):={A(n)}/{g_i(n)}, B(n):={B(n)}/{g_i(n-h_i)}$;
\item Return $(A,B,\{(g_i(n),h_{i}),i=1,\dots,s\})$. 
\end{enumerate} 
\end{minipage}
}
\end{center}
\begin{example}${\sf CompactGPF}(n,n-N)=(1,1,\{(n,N)\})$.
\end{example}
Note that the input is an ordered pair $(P,Q)$ and not a rational function $P/Q$. The 
output of the algorithm changes if $(P,Q)$ is replaced by $(FP,FQ)$ for $F\in\Q[n]$. This will be necessary for our treatment of rational solutions below. On the other hand, the output $A$, $B$, and~$g_i$'s also satisfy~(\ref{gp-ext}) whenever $P$ and~$Q$ have no common factor, so that the Gosper-Petkov\v{s}ek normal form of a rational function in~$\Q(n)$ given in reduced form~$P/Q$ is obtained by ${\sf CompactGPF}(P,Q)$.

As an outcome of this algorithm, the rational function $C(n)/C(n+j)$ ($j=1,2,\dots$) is easily obtained as
\begin{equation}\label{Cnplus1overCn}
\frac{C(n)}{C(n+j)}=
\prod_{i=1}^{s}\frac{g_{i}(n+j-1-h_{i})\dotsm g_{i}(n-h_{i})}{g_{i}(n+j-1)\dotsm g_{i}(n)}.
\end{equation}
For large~$N$ and~$j=\bigO(1)$, it has ``small'' numerator and denominator of degrees bounded by $j$ times those of~$P$ and~$Q$. This equation for $j=1$ is a homogeneous LRE that plays the role of a compact representation of~$C$. The initial value~$C(0)$ (more generally~$C(k)$ where~$k=\bigO(1)$) has size~$\bigO(N\log N)$ and can be computed by Thm.~\ref{thm:reccomplexity} within~$\bigO(I(N\log N)\log N)$ bit operations. In the next sections, we use this to design ``compact'' variants of Abramov's and Gosper's algorithms.

\begin{proposition} \label{logH}
Algorithm \textsf{CompactGPF} is correct. For $(P,Q)$ with rational coefficients, it has deterministic polynomial bit complexity in the bit size of $(P,Q)$.
\end{proposition}
\begin{proof}
The correctness is that of the classical algorithm since the only difference is that we do not expand~$C$.
Step~1 is dealt with by a deterministic algorithm due to Loos~\cite{Loos83} (cf.~\cite{Gerhard04,GeGiStZi03} for faster probabilistic algorithms). Step~3 is performed at most~$\deg P\deg Q$ times, and each step is polynomial by the classical algorithms as found in~\cite{GaGe99}.
\end{proof}
      
\subsection{Compact Rational Solutions}
We now consider rational solutions of the LRE $\mathcal{R}u(n)=f(n)$, with
$f$ a polynomial in $\Q[n]$.

Our starting point is the following result of Abramov~\cite{Abramov95c}.
\begin{lemma}[Abramov]\label{Abramov-lemma}
The polynomial $C(n)$ of the\\ Gosper-Petkov\v{s}ek form of $(a_r(n-r+1),a_0(n))$ 
is a multiple of the denominator of all rational solutions of~$\mathcal{R}u(n)=f(n)$. 
\end{lemma}
Abramov's algorithm first computes $C(n)$, then performs 
the change of variable $u(n)={v(n)}/{C(n)}$, leading to
\begin{equation} \label{abrpol}
a_r(n) \frac{v(n+r)}{C(n+r)} + \cdots + a_0(n) \frac{v(n)}{C(n)} =
f(n),
\end{equation} 
whose polynomial solutions $v(n)$ are then sought.

In the homogeneous case ($f(n)=0$), using~\eqref{Cnplus1overCn} reduces this equation to an equation of polynomial size. This is described in Algorithm {\sf HomCompactRatSols} (see Figure).
\begin{figure}
\begin{center}
\fbox{
\begin{minipage}{8cm}
\begin{center}
\textsf{Algorithm HomCompactRatSols}\\
\end{center}
{\bf Input:} a homogeneous LRE $\mathcal{R}u(n)=0$.\\
{\bf Output:} a basis of its rational solutions in compact form
\begin{enumerate}
\item $(A,B,C):=\textsf{CompactGPF}(a_r(n-r+1),a_0(n))$; 
\item Normalize~$C(n)\mathcal{R}(v(n)/C(n))$ using~\eqref{Cnplus1overCn} and denote the result $\mathcal{T}v(n)$;
\item Compute a basis $\mathcal{B}$ of the polynomial solutions of $\mathcal{T}v(n)=0$;
\item Return $\{p(n)/C(n)\mid p(n)\in\mathcal{B}\}$. 
\end{enumerate} 
\end{minipage}
}
\end{center}
\end{figure}
In Step~2, the ``Normalize'' operation consists in expanding~$C(n)/C(n+j)$ using~\eqref{Cnplus1overCn} and taking the numerator of the resulting expression. Also, if necessary, we change $n$ into~$n+\alpha$ with~$C(\alpha)\neq0$, so that~0 is not a singular point 
in Step~3. This can be detected and changed at a cost of~$\bigO(\I(N\log N)\log N)$ operations. In Step~4, the output is given by the compact forms of the numerators and $C$ is given by the output of~\textsf{CompactGPF}.

In the nonhomogeneous case, reducing~\eqref{abrpol} to the same denominator would lead to an equation whose right-hand side has a potentially exponential degree. Instead, we consider the homogeneous operator~$\mathcal{S}=(f(n)S_n-f(n+1))\mathcal{R}$, whose bit size is polynomial in that of~$\mathcal{R}u(n)=f(n)$ and that can be treated by the algorithm above. If~$u_n$ is a rational solution of~$\mathcal{S}$, then~$w_n=\mathcal{R}u_n$ is a rational solution of~$f(n)w_{n+1}=f(n+1)w_n$. This implies that $w_n=\lambda f(n)$ for all~$n$ larger than the largest root of~$f$ and since $w_n$ is rational, also for all other values of~$n$. Thus fixing~$\lambda$ so that $\mathcal{R}u(k)=f(k)$ for any $k$ such that~$f(k)\neq0$ concludes the computation. This is the basis of the following algorithm.
\begin{center}
\fbox{
\begin{minipage}{8cm}
\begin{center}
\textsf{Algorithm NonhomCompactRatSols}\\
\end{center}
{\bf Input:} a LRE $\mathcal{R}u(n)=f(n)$, with $f\neq0$.\\
{\bf Output:} a particular rational solution~$p$ and a basis~$(b_1,\dots,b_d)$ of rational solutions of~$\mathcal{R}u$ in compact form
\begin{enumerate}
	\item $W:=\textsf{HomCompactRatSols}((f(n)S_n-f(n+1))\mathcal{R});$
	\item Find~$k\in\N$ such that $f(k)\neq0$;
	\item Write $\mathcal{R}(\sum_{w\in W}{\xi_ww(k)})=:\mathcal{U}(\xi)$ for an unknown $\xi=(\xi_w)_{w\in W}$ and solve $\mathcal{U}(\xi)=0$ for a basis $(\mu^{(1)},\dots,\mu^{(d)})$ of its solution space and $\mathcal{U}(\xi)=f(k)$ for a particular solution~$\lambda$;
	\item Return $p:=\sum_{w\in W}\lambda_ww(n)$ and the~$b_i$'s given by $b_i:=\sum_{w\in W}\mu^{(i)}_ww(n)$.
\end{enumerate} 
\end{minipage}
}
\end{center}
In Step~2, just iterating $k=0,1,\dots$ till a point where $f$ is found to be nonzero is sufficient for our purpose. 
If~$N$ is a bound on the degree of the numerators and denominator computed in Step~1, 
then the values of the~$w(k)$'s in Step~3 have size~$\bigO(N\log N)$ and can be computed by binary splitting. 
From there, it follows that the affine equation in Step~3 has coefficients of size~$\bigO(N\log N)$, which is then also a bound on the size of its solutions. These solutions can be computed in the form of a point and a basis of a vector space within~$\bigO(\I(N\log N)\log N)$ bit operations by standard linear algebra. The same complexity is sufficient for the products of initial conditions in Step~4.

The results of this section are summarized as follows.
\begin{theorem}
Let~$N$ be the sum of the largest nonnegative integer root of the indicial polynomial of~$\mathcal R$ at infinity and the degree of the polynomial~$C(n)$ of~\eqref{gp} with~$P(n)=a_{r}(n-r+1)$ and $Q(n)=a_{0}(n)$. 
The dimension of the affine space of rational solutions of~$\mathcal{R}u(n)=f(n)$ can be computed probabilistically using~$\bigOsoft(\M(\sqrt{N})\I(\log N))$ bit operations. A compact representation of the solutions can be computed deterministically in~$\bigO(\I(N\log N)\log N)$ bit operations.
\end{theorem}
\begin{proof}
The largest integer root of the indicial polynomial of~$\mathcal R$ at infinity is a bound on the valuations of power series solutions of~$\mathcal{R}u=0$ at infinity, including the valuation of~$v(n)/C(n)$. Adding the degree of~$C$ gives the announced bound on the degree of polynomial~$v$'s. From there, the theorem follows from Cor.~\ref{cor:finitesuppsols}.
\end{proof}

\subsection{A Compact Gosper Algorithm}
Given a hypergeometric term~$t(n)$, i.e., such that~$t(n+1)/t(n)=:r(n)\in\Q(n)$, 
Gosper's algorithm~\cite{Gosper78} finds its indefinite hypergeometric sum, if it exists. Such a sum is necessarily of the form~$u(n)t(n)$ for some $u(n)\in\Q(n)$. Thus, the problem is reduced to finding rational solutions of 
$u(n+1)r(n)-u(n)=1$.
This can be solved by~\textsf{NonhomCompactRatSols}.
A further optimization is present in Gosper's algorithm: if~$r(n)=P(n)/Q(n)$ in reduced form, the polynomial~$B(n)$ of~\eqref{gp} satisfies~\eqref{gp-ext}, so that it divides the numerator of~$u(n+1)$. (This can be generalized to detect factors of numerators in arbitrary
LRE's). This does not affect the expression of the complexity result, which is as follows.
\begin{theorem}
Let $t(n)$ be a hypergeometric term such that $t(n+1)/t(n)=:P(n)/Q(n)\in\Q(n)$, with $\gcd(P,Q)=1$. 
Let $N$ be a bound on the degree of~$C$ in~\eqref{gp} and on the largest positive integer root of the indicial polynomial of~$P(n)S_n-Q(n)$ at infinity. Then the existence of an indefinite hypergeometric sum of~$t(n)$ can be determined by a probabilistic algorithm using~$\bigOsoft(\M(\sqrt{N})I(\log N))$ bit operations, a compact representation of it can be computed deterministically using~$\bigO(\I(N\log N)\log N)$ bit operations.
\end{theorem}
Note that in the special case of rational summation (i.e., $t(n)\in\Q(n)$), it is actually possible to decide the existence of a rational sum in only \emph{polynomial} complexity, see~\cite{GeGiStZi03}.

\section{Definite Hypergeometric Sums}\label{zeilberger}

A bivariate hypergeometric term $t(n,m)$ is such that both $t(n+1,m)/t(n,m)$ and $t(n,m+1)/t(n,m)$ belong to~$\Q(n,m)$. Given such a term, Zeilberger's algorithm~\cite{Zeilberger91b} computes a LRE satisfied by
$T(m)=\sum_{n}t(n,m)$.
The idea is to synthesize a
{\em telescoping recurrence}, i.e., a rational function~$u(n,m)$ and a linear 
operator $P(m,S_m)$ such that
\[
(S_n-1)  u(n,m)  t(n,m) = P(m,S_m)  t(n,m).
\]
Indeed, summing over $n$ and granted boundary conditions known as ``natural boundaries'', we obtain $P(m,S_m) T(m) =0$. If~$P$ was known, then Gosper's algorithm would find the left-hand side. This is the basis of Zeilberger's algorithm (see Figure).
\begin{figure}
\begin{center}
\fbox{
\begin{minipage}{8cm}
\begin{center}
\textsf{Zeilberger's Algorithm}\\
\end{center}
{\bf Input:} two functions $\frac{t(n+1,m)}{t(n,m)}$ and $\frac{t(n,m+1)}{t(n,m)}$ in~$\Q(n,m)$.\\
{\bf Output:} a  LRE $\sum_{i=0}^r{\lambda_i(m)S_m^i}(\sum_{n}t(n,m))=0$.\\[1mm]
For $r=0,1,2,\dots$ do
\begin{enumerate}
	\item Construct the equation~$(E_r)$
\begin{small}	\begin{equation*}
	u(n+1,m)\frac{t(n+1,m)}{t(n,m)}-u(n,m)=\sum_{i=0}^r\lambda_i(m)\frac{t(n,m+i)}{t(n,m)},
	\end{equation*}
\end{small}
	\item Find if there exist~$\lambda_i$'s in~$\Q(m)$ so that~$(E_r)$ admits a solution~$u(n,m)\in\Q(n,m)$;
	\item If so, compute and return them; otherwise proceed to the next~$r$.
\end{enumerate} 
\end{minipage}
}
\end{center}
\end{figure}
Termination is guaranteed only if such a LRE exists. This occurs in the so-called ``proper-hypergeometric'' case~\cite{WiZe92b} and a general criterion has been given by Abramov~\cite{Abramov02b}.

Note that knowing~$u$ permits to check the output operator~$P$ by simple rational function manipulations, which is why the rational function~$u$ is called ``certificate'' in~\cite{PeWiZe96}.

Zeilberger's algorithm is based on a refinement of Gosper's algorithm for Steps~2 and~3. It reduces the computation in Step~2 to solving a system that is linear simultaneously in the~$\lambda_i$'s and in another set of~$N+1$ variables, where~$N$ is potentially exponential in the bit size of $(E_r)$, see e.g.~\cite[\S6.3]{PeWiZe96}. An equivalent linear system in a small number of variables can be computed by Algorithm \textsf{Small Linear System} (see Figure).
\begin{figure}
\begin{center}
\fbox{
\begin{minipage}{8cm}
\begin{center}
\textsf{Small Linear System}\\
\end{center}
{\bf Input:} the equation $(E_r)$ from Zeilberger's algorithm.\\
{\bf Output:} an equivalent system linear in the~$\lambda_i$.
\begin{enumerate}
	\item Compute~$\mathcal{R}u(n)=f(n)$, the numerator of~$(E_r)$;
	\item Compute a multiple~$C(n)$ of the denominator of its rational solutions and a bound~$N$ on the degree in~$n$ of their numerators;
	\item Compute~$\mathcal{S}v(n)$, the numerator of \\$C(n)(f(n)S_n-f(n+1))\mathcal{R}(v/C)(n)$;
	\item Compute~$(\mathcal{T},\mathcal{E}):=\textsf{RecToRec}(\mathcal{S})$; set~$\mathcal{E}:=\mathcal{E}\cup\{\mathcal{R}(v/C)(0)=f(0)\}$; let $s$ be the order of~$\mathcal{T}$;
	\item Compute the value~$(c_{N+1},\dots,c_{N+s})=:V$ for a nonzero sequence solution of~$\textsf{RecToRec}(C\mathcal{R}(v/C))$;
	\item Compute the value~$W:=(d_{N+1},\dots,d_{N+s})$ for an arbitrary sequence solution of~$\mathcal{T}$ obeying~$\mathcal{E}$; $W$ is of the form~$W^\star+\sum_{i=0}^r\lambda_iW_i$, only~$W^\star$ depends on the initial conditions;
	\item The system $(\Sigma):=(\mu V+\sum_{i=0}^r\lambda_iW_i=0)$ is simultaneously linear in the~$\lambda_i$'s and~$\mu$.
\end{enumerate}
\end{minipage}
}
\end{center}  
\end{figure}
The important point is linearity: not all solutions of~$\mathcal T$ are linear in the~$\lambda_i$'s, but this property is ensured when the initial conditions satisfy~$\mathcal E$. Indeed, in Step~2, by Lemma~\ref{Abramov-lemma}, $C$ does not depend on the~$\lambda_i$'s. Then, by induction on $n$, starting from $\mathcal{R}(v/C)(0)=f(0)$, the factor~$f(n)$ of the leading coefficient in~$\mathcal{S}$ cancels out and thus the solution~$v(n)$ is linear in the~$\lambda_i$. This property is then preserved by the linearity of \textsf{RecToRec}. The final system~$(\Sigma)$ has solutions if and only if $(E_r)$ has rational solutions.

The description of \textsf{Small Linear System} is geared towards the use of compact representations and matrix factorials in intermediate steps. This is straightforward for Steps~1--5. In Step~6, we cannot make direct use of the factorial of the  matrix associated to~$\mathcal{T}$: this matrix involves the~$\lambda_i$'s rationally and its factorial has too large a size for our target complexity. Instead, we exploit the linearity in the~$\lambda_i$'s by constructing the vector~$W$ using matrix factorials for $\lambda$ a vector of 0's with a~1 in $i$th position for~$i=0,\dots,r$ and setting the initial condition to~0, which gives the coefficients~$W_i$.

From there we derive our compact version of Zeilberger's algorithm given in \textsf{Compact Zeilberger Algorithm}.
\begin{figure}
\begin{center}
\fbox{
\begin{minipage}{8cm}
\begin{center}
\textsf{Compact Zeilberger Algorithm}\\
\end{center}
{\bf Input:} two functions $\frac{t(n+1,m)}{t(n,m)}$ and $\frac{t(n,m+1)}{t(n,m)}$ in~$\Q(n,m)$.\\
{\bf Output:} a  LRE $\sum_{i=0}^r{\lambda_i(m)S_m^i}(\sum_{n}t(n,m))=0$.\\[1mm]
For $r=0,1,2,\dots$ do
\begin{enumerate}
	\item Take a random~$m_0\in\Q$ and construct~$(E_r)$ with~$m=m_0$;
	\item Apply~\textsf{Small Linear System} to this equation;
	\item Find if there exist nonzero solutions to this system;
	\item If not, proceed to the next~$r$;
	\item Otherwise, construct~$(E_r)$, apply~\textsf{Small Linear System} and return its solutions.
	If it does not have nonzero rational solutions, go to Step~1.
\end{enumerate}
\end{minipage}
}
\end{center}  
\end{figure}
In Step~2, the whole construction can be performed by matrix factorials with integer entries, within the complexities of Thm.~\ref{thm:reccomplexity}. If a rational solution $(\lambda_i(m))$ exists, then the system~$(\Sigma)$ has the corresponding~$(\lambda_i(m_0))$ for solutions. Thus if $(\Sigma)$ does not have a nonzero solution, $(E_r)$ does not have a rational one. This gives a fast probabilistic test.
Then, in Step~5, the algorithm is used again with matrices that are polynomial in the variable~$m$. In that case, the system~$(\Sigma)$ can be computed by binary splitting with~$\bigOsoft(\M({N})\log N)$ \emph{arithmetic} operations. The final system has coefficients of degree~$\bigO(N)$ with coefficients of bit size~$\bigO(N\log N)$ each and this is also the size of the~$\lambda_i$'s to be found. At the same time, we find~$\mu$,  which gives us 
a compact representation of the certificate.

An optimization is obtained by using the values of the $\lambda_i(m_0)$'s to compute the value~$N'$ of the degree of the corresponding sequence. With high probability this is the actual degree in~$n$ of the numerator of~$u(n,m)$, which can be much smaller than~$N$, thus saving a lot of computation in Step~5.

The following theorem summarizes this section.
\begin{theorem}\label{thm:zeilberger}Let~$t(n,m)$ be hypergeometric over~$\Q$.
	Let~$N$ be the maximal number of variables in the linear system solved in the classical version of Zeilberger's algorithm. Then it is possible to detect probabilistically that this system does not have any nonzero solution in~$\bigOsoft(\M(\sqrt{N})\I(\log N))$ bit operations. If it does have a solution, it is possible to compute the corresponding~$\lambda_i$'s of degree~$\bigO(N)$ and total bit size $\bigO(N^2\log N)$, as well as a compact representation of the certificate, in $\bigOsoft(\M(N)\I(N\log N))$ bit operations.
\end{theorem}
For the sake of comparison,
a crude analysis by unrolling the triangular system of dimension $N+r+2$ and taking into account coefficient growth leads to a~$\bigOsoft(N^4)$ bit complexity estimate for the classical algorithm, which can be reduced to~$\bigOsoft(N^3)$ by using the binomial basis.

\section{Experimental Results}\label{experiments} 
\subsection{Rational Solutions}
We consider two families of linear recurrences:
\begin{small}
\begin{multline*}
	2n ( N-n)  ( -4N-3nN+6+3{n}^{2}+8n ) u ( n
	 )\\ - ( n+1 )  ( -3nN+2N+3{n}^{2}-4n-4 )  ( n+1-
	N ) u ( n+1 ) \\
	+ ( n+2 )  ( -3nN-N+3{n}^{2}+2n+1
	 )  ( n+2-N ) u ( n+2 ) =0,\\
	2n ( n-2N )  ( n-N )  ( {n}^{2}-3nN+3n+2{N}^{2}-3
	N+2 ) u ( n ) \\- ( n+1 )  ( n+1-2N )  ( n+1
	-N )  ( 3{n}^{2}+6n-9nN+6{N}^{2}-4N ) u ( n+1 ) \\+
	 ( n+2 )  ( n+2-2N )  ( n+2-N )  ( {n}^{2}+n-3
	nN+2{N}^{2} ) u ( n+2 ) =0.
\end{multline*}
\end{small}
The first one ($R_1$) does not have any rational solution, while the second one ($R_2$) has $1/(n(n-2N))$ as a solution. In both cases, when~$N$ is a large integer, a large dispersion has to be considered.
In Table~\ref{table-abramov}, we give a comparison of the timings\footnote{All our tests have been run on an Intel Xeon at 3.6GHz.} obtained by our Maple prototype (denoted Compact) and that of the command \texttt{ratpolysols} of Maple's package \texttt{LREtools} (denoted Classical). This table illustrates the ``nonexponential'' character of the compact versions of the algorithms. In the first case, both output are identical (no solution). In the second case, however, we return a compact representation of the output. For instance, with~$N=2^{100}$ we get (in 0.04s) the denominator $(n(n-2^{100})(n-2^{101}))$ (in expanded form) and for the numerator the recurrence 
\begin{multline*}
(1-k^2)c_k+(2^{100}+Ak-k^2)c_{k+1}\\
+(k^2-2k-B)c_{k+2}+(k^2-Ck+D)c_{k+3}=0,
\end{multline*}
satisfied by its coefficients in the binomial basis, together with initial conditions~$c_0=-2^{100}$, $c_1=1$, where the coefficients~$A,\dots,D$ are 200 bit long integers.
\begin{table}[t] 
  \begin{center} 
\begin{tabular}{|c|c|c|c|c|} 
\hline
& \multicolumn{2}{|c|}{Classical}&\multicolumn{2}{|c|}{Compact}   
\\ \cline{2-5}
$N$ 
&$R_1$&$R_2$&$R_1$&$R_2$\\
\hline 
 $2^7$  & 5.4    & 0.1    & 0.044&0.019\\
 $2^8$  & 52.8   & 0.1    & 0.046&0.019\\
 $2^9$  & 518.0  & 0.2    & 0.048&0.021\\
$2^{10}$&$>$10000&1.0     & 0.048&0.021\\
$2^{11}$&        &6.2     & 0.049&0.021\\
$2^{12}$&        &46.4    & 0.051&0.022\\
$2^{13}$&        &362.0   & 0.052&0.023\\
$2^{14}$&        &2860.   & 0.053&0.023\\
$2^{15}$&        &$>$10000& 0.055&0.024\\[1mm]
$2^{40}$&        &        &0.083 &0.037\\
\hline  
\end{tabular} 
\end{center} 
\caption{Timings (in sec.) for classical and compact versions of Abramov's algorithm\label{table-abramov}}
\end{table}

\subsection{Definite Hypergeometric Summation}
We consider the following family of hypergeometric terms:
\[t(n,m)=\binom{2n+m+N}{N}\binom{2m}{2n}\binom{m}{n}.\]
For~$N\in\N$, the sum~$\sum_n{t(n,m)}$ satisfies a third-order homogeneous LRE.
When Zeilberger's algorithm is executed on this term, 
the bound it has to use on the degrees of numerators of rational solutions of the equation~$(E_r)$ is $N+3(r-1)$. This plays the r\^ole of a ``large''~$N$ and makes it possible to exhibit the complexity behaviour of the algorithms.

\begin{table}[t] 
  \begin{center} 
\begin{tabular}{|c||c|c|c|c|} 
\hline 
& \multicolumn{4}{|c|}{Classical}   
\\ \cline{2-5}
$N$ 
&$r=0$&$r=1$&$r=2$&$r=3$\\
\hline 
16 & 0.1  &  0.2 & 0.3 & 0.6\\ 
32 & 0.3 & 0.7 & 1.5 & 3.4  \\ 
64 & 2.9 & 6.8 & 12.  & 34.3\\ 
128 & 43.9   &  131.0  & 276.4 & 1202.6 \\ 
256 & 1793.4  & $>2$Gb &&\\
\hline  
\hline 
& \multicolumn{4}{|c|}{Compact, random $m$}\\   
$N$ 
&$r=0$&$r=1$&$r=2$&$r=3$\\
\hline 
16 & 0.1 & 0.3 & 0.9 & 2.5  \\
32 & 0.1 & 0.5 & 1.4 & 5.1  \\
64 & 0.2 & 0.7 & 2.6 & 7.3  \\
128 & 0.3 & 1.5 & 5.0 & 15.2  \\
256 & 0.5 & 2.7 & 11.3 & 35.5  \\
512 & 1.0 & 6.3 & 27.8 & 106.2  \\
1024 & 2.2 & 15.7 & 72.7 & 240.1  \\
\hline  
\end{tabular} 
\end{center} 
\caption{Timings (in sec.) for classical and compact versions of Zeilberger's algorithm\label{timings}}
\end{table} 

In Table~\ref{timings}, we give a comparison of the timings obtained by our prototype implementation in Maple (denoted ``Compact'') and those obtained by Maple's \texttt{Zeilberger} command in the package \texttt{SumTools:-Hypergeometric} (denoted ``Classical'').
The indication ``$>2$Gb'' means that the computation had to be stopped after two gigabytes of memory had been exhausted. The first part of the table (Classical) suggests that the implementation does not behave well for large~$N$: the observed behaviour is exponential instead of polynomial. Even then, it is still much better than our implementation. Indeed, we have implemented only the case with rational values of~$m$ and for small~$N$ it often takes longer for our implementation to compute the result with this value than for the classical method to find the result with a formal~$m$. However, things change as $N$ gets larger: the predicted behaviour is well observed. When $N$ is multiplied by~2, the time is multiplied by slightly more than~2. Had we implemented the baby-step/giant-step version of binary splitting, the timings in the columns for random~$m$ would have been much better, since the time should be multiplied by slightly more than~$\sqrt{2}$ from one line to the next.
Our experiments with symbolic~$m$ show that so far, our complexity result is more of a theoretical nature: although the degrees of the coefficients of the equations grow like~$\bigO(N)$, the constant in front of the~$\bigO$ term is about~18 in this example, and a massive cancellation takes place in the final linear solving. The result has degrees that also grow like~$\bigO(N)$, but with a much smaller constant, so that a direct resolution in~$\bigOsoft(N^4)$ is much faster in this range than our~$\bigOsoft(N^2)$.

\section*{Acknowledgements}
This work was supported in part by the French National Agency for Research (ANR Gecko). Comments of the referees on the first version of this article have been very useful.

\bibliographystyle{abbrv} 
\def\gathen#1{{#1}}

\end{document}